\documentstyle[prl,aps,twocolumn,epsf]{revtex}

\begin{document}

\draft

\title{Sound attenuation in the superconducting amorphous alloy ZrTiCuNiBe}

\author{E. V. Bezuglyi, A. L. Gaiduk, and V. D. Fil}

\address{B. Verkin Institute for Low Temperature Physics and Engineering,
National Academy of Sciences of the Ukraine, 310164 Kharkov, Ukraine\\E-mail:
fil@ilt.kharkov.ua}

\author{W. L. Johnson}
\address{California Institute of Technology, Pasadena, CA 91125, USA}

\author{G. Bruls, B. L\"{u}thi, B. Wolf, and S. V. Zherlitsyn}

\address{Physikalisches Institut, Universit\"{a}t Frankfurt, 60054
Frankfurt, Germany}

\date{Submitted May 25, 1999}

\maketitle

\begin{abstract}
The superconducting energy gap and the parameter $\eta$ of the intensity of
electron scattering at two-level systems in amorphous ZrTiCuNiBe are
determined from the results of measurements of sound attenuation. The
mechanism of adiabatic renormalization of the amplitude of coherent tunneling
is used for a quantitative description of the peculiarities of sound
absorption in the vicinity of $T_c$.
\end{abstract}

\narrowtext \vspace{0.2cm}

Preliminary measurements of the velocity $v$ and absorption $\Gamma$ of sound
in the amorphous alloy Zr$_{41.2}$Ti$_{13.8}$Cu$_{12.5}$Ni$_{10}$Be$_{22.5}$
have revealed \cite{1} an interesting peculiarity in its behavior in the
vicinity of the superconducting transition temperature $T_c$. It was found
that the value of $T_{cm} \simeq 0.9$ K determined from magnetic
measurements exceeds the transition temperature $T_c \simeq 0.83$ K at which
a difference is observed in the velocities of sound in superconducting ($s$)
and normal ($n$) phases. It was proposed that such a behavior may be due to
magnetic depairing leading to the gapless $s$-phase in the temperature
interval $T_{cm}$--$T_c$. Since our earlier measurements \cite{1} were made
in a temperature interval limited from below ($T \agt 0.4$ K), it should be
interesting to study the behavior of $\Gamma$ at lower temperatures. In view
of a close analogy in the behavior of electron acoustic absorption
coefficient in superconductors and the relaxation absorption of sound by
two-level systems (TLS) in the $s$-phase, one could expect a nonexponential
drop in $\Gamma_s(T)$ in the low-temperature ``tail'', or an exponential
decrease but with a much smaller gap if the magnetic depairing effects are
significant. It is shown in the present work that the $\Gamma_s(T)$
dependence at low temperatures can be described quite correctly by the tunnel
model (TM) \cite{2} in the framework of the standard BCS approximation, which
rules out the gapless phase hypothesis. The refined value of the transition
temperature for the bulk of the sample was found to be close to $T_c$, while
the value of $T_{cm}$ determined from magnetic measurements is apparently
associated with the surface phase. However, the behavior of $\Gamma$ in the
vicinity of $T_c$ does not conform to the standard TM: the drop in
$\Gamma_s(T)$ below $T_c$ begins much later than what is predicted by TM.
Besides, in a certain temperature interval below $T_c$, the absorption is
slightly higher than $\Gamma_n(T)$. Such effects were observed earlier in the
amorphous alloy Pd$_{30}$Zr$_{70}$ and were explained qualitatively by the
electron renormalization of the parameter of TLS interaction with an elastic
wave \cite{3}. But possible mechanisms of such a renormalization were not
discussed and quantitative estimates of its magnitude were not obtained in
Ref.~3. We shall use the mechanism of adiabatic renormalization of the
amplitude of coherent tunneling \cite{4} to explain the peculiarities of
sound absorption in the vicinity of $T_c$. This approach gives an acceptable
quantitative description of the dependence $\Gamma_s(T)$.

The basic postulate of TM, which was confirmed irrefutably in experiments
(see the review by Hunklinger and Raychaudhuri \cite{2}), is the assumption
that glasses have double-well potentials with a tunneling bond between the
wells whose density of states $\overline{p}$ is constant in the space of the
parameters $\xi$ and $\ln \Delta_0$ ($\xi$ is the asymmetry of a double-well
potential and $\Delta_0$ the amplitude of coherent tunneling). The response
of the TLS system to an external perturbation is determined by the average
over the TLS ensemble. For the sake of convenience, averaging is usually
carried out by using new variables $E = \sqrt{\xi^2 + \Delta_0^2}$ and $u =
\Delta_0/E$ in which the density of states of TLS is independent of $E$:
\begin{equation}
g(E,u)= {\overline{p} \over u\sqrt{1-u^2}} \equiv {g(u)}.
\end{equation}

Under the conditions of the experiment ($\omega \ll T$, $\omega$ being the
frequency of acoustic vibrations), the attenuation of sound associated with
the TLS is determined by the relaxation mechanism  and is described by the
following standard expression \cite{2}:
\begin {equation}
\left({\Gamma v \over \omega} \right)_{rel}\!\!\! =
\!\!\int_0^{E_g/T} \!\!\!\!\!\!\!\!\!\!\!\!{d\varepsilon \over
\cosh^2(\varepsilon/2)}\!\! \int_0^{1}\!\!\!\!\!C
g(u)(\!1\!-\!u^2) {\omega \nu \over \omega^2 +\nu^2} du.
\end {equation}
Here $\varepsilon = E/T$, $E_g \gg T$ is the limiting energy, and $\nu$ the
relaxation frequency. In Eq.~(2) and below, we have used the system of energy
units ($\hbar = k = 1$). The order of magnitude of the TLS contribution to
the velocity  and attenuation of sound is determined by the parameter $C$. In
the standard TM \cite{2}, this quantity is a constant: $C = C_0 = \overline{p}
\gamma^2/(\rho v^{2})$ ($\gamma=1/2(\partial{\xi}/ \partial e)$ is the
deformation potential, $e$ the deformation, and $\rho$ the density), although
a number of experimental facts can be explained only by assuming that $C$
depends on $E, u$ or $T$.

The TLS relaxation is due to their interaction with electrons as well as
phonons, although the contribution of the latter can be disregarded for $T
\lesssim$ 1 K. The intensity of interaction of TLS with electrons is
determined by the dimensionless parameter $\eta=2n_0\sqrt{\overline
{V_{kk'}^2}}$, where $n_0$ is the density of electron states at the Fermi
level and $\overline{V_{kk'}^2}$ is the square of the matrix element of
electron scattering at TLS from the state $\bbox{k}$ to the state $\bbox{k}'$,
averaged over the Fermi surface.

In the standard TM, the interaction of TLS with electrons is considered
within the framework of the perturbation theory in parameter $\eta^2$, which
does not change the system of energy levels \cite{5}. The entire distinction
between a metallic glass and an amorphous insulator is reflected just in the
emergence of a new relaxation channel with a characteristic rate
\begin {equation}
\nu={\pi \eta^2\over 2} Tu^{2}J(\varepsilon).
\end {equation}

In the $n$-phase, $J(\varepsilon)=J_{n}(\varepsilon) = (\varepsilon/2)
\coth(\varepsilon/2)$, $\nu \approx \eta^{2} Tu^{2}$. As long as $\omega \ll
T$, there always exist TLS with $\nu_{\rm opt} \sim \omega$, and the
absorption~(2) is practically independent of temperature (``plateau''
region).

In the $s$-state, we must use instead of $J_n(\varepsilon)$ the function
$J_s(\varepsilon,\Delta)$ ($\Delta = \Delta_s/T$, $\Delta_s$ is the
superconducting energy gap) \cite{6}:
\[
J_s(\varepsilon,\Delta)= {1\over 2} \int_\Delta^\infty d\varepsilon'
\sqrt{\varepsilon'^2 - \Delta^2} \left({\varepsilon'(\varepsilon'-
\varepsilon) - \Delta^2 \over \sqrt{(\varepsilon'-\varepsilon)^2-\Delta^2}}
\times\right.
\]
\begin {equation}
\left.\times{f(\varepsilon'\!\!\!-\! \varepsilon) \over
f(-\varepsilon)} \Theta((\varepsilon'\!\!\!-\!\varepsilon)^2\!-
\!\Delta^2)\; \mbox{sign} (\varepsilon'\!\!\!- \!\varepsilon)
+(\varepsilon\!\!\rightarrow\!\! -\varepsilon) \right)
\end {equation}
where $f(x)$ is the Fermi function and $\Theta(x)$ is the step
$\Theta$-function. The function $J_s(\varepsilon, \Delta)$ is frequently
encountered in the theory of kinetic properties of superconductors. It has a
discontinuity at $\varepsilon = 2\Delta$, while for $\varepsilon\ll
2\Delta\;\; J_s(\varepsilon,\Delta)\rightarrow 2f(\Delta)$. As a result of a
rapid drop in the value of $J_s$ below $T_c$, the maximum relaxation rate ($u
= 1$) becomes less than $\omega$ starting from a certain temperature and
$\Gamma_s(T)$ ``freezes''.

A nonperturbative analysis \cite{4,7} revealed a more complicated pattern.
Even at $T =0$, the initial coherent tunneling amplitude $\Delta_0$ in the
$n$-phase is renormalized as the adiabatic part of the interaction of TLS
with electrons is taken into consideration:
\begin {equation}
\Delta_0^{\ast} \approx \Delta_0 \left( {\Delta_0 \over \omega_0}
\right)^{\displaystyle{\eta^2 \over 4-\eta^2}},
\end {equation}
where $\omega_0$ is the energy of the order of Debye energy.

For $T \neq 0$, the TLS ensemble can be divided conditionally into three
intervals according their position on the $E$-scale in the $n$-state.

1) The coherent tunneling region $E^{\ast}\!=\!\sqrt{\xi{^2}\!+\!
\Delta_0^{\ast 2}} > T$.

2) The region ${E^{\ast}}<T<4\widetilde E/(\pi \eta^2)$ of incoherent
tunneling with an amplitude $\widetilde{\Delta} = \Delta_0 \left( 2\pi
T/\omega_0\right)^{\eta^2/4}$ and the energy splitting $\tilde{E} =
\sqrt{\xi^2 + \tilde{\Delta}^2}$. Going over to renormalized variables $E^*$
and $\tilde{E}$ as well as to $u^*$ and $\tilde{u}$ during averaging in each
of the regions~1 and 2, relations (2) and (3) remain valid.

3) The low-energy TLS $T > 4 \tilde{E}/(\pi \eta^2)$. In this region also,
the tunneling is incoherent and has an amplitude $\tilde{\Delta}$. However,
the factor $(1 -\tilde{u}^2)$ vanishes from Eq.~(2) -- a reflection of the
fact that the incoherent transitions between broadened levels occur with a
variation of energy even in the symmetric case. The relaxation frequency
\begin {equation}
\nu_3 \approx {2\over\pi\eta^2}T\widetilde{u}^2 \left(
{\widetilde{\varepsilon}\over T}\right)^2 {1\over J(\varepsilon)}.
\end {equation}
also changes in region 3.

It would appear that as a result of a decrease in $\nu_3$ for small
$\tilde{\varepsilon}$ [Eq.~(6)], the contribution to $\Gamma$ from the part
of TLS with $\tilde{E} < \sqrt{\omega T}$ must decrease. However, this
decrease is compensated by an increase in the contribution from symmetric
TLS, and the partial contribution from region~3 to $\Gamma$ remains
practically the same as that calculated in the standard TM. The contribution
from region~2 also remains unchanged. Only the contribution from coherently
tunneling TLS (region~1) undergoes significant variation. Upon a transition
to the variables $E^*$ and $u^*$, the density of states $g(u^*)$ (1) is
renormalized as a result of a nonlinear relation~(5) between $\Delta_0^*$ and
$\Delta_0$, and acquires an additional factor $(1 - \eta^2/4)$. The parameter
$C$ is also renormalized accordingly. If the boundary between regions 1 and 2
is located at $E^* \sim T$, the resulting value of $\Gamma$ in the $n$-phase
decreases in spite of the fact that the denominator in~(2) decreases the
contribution from the high-energy TLS. Below $T_c$, the nonlinear
relation~(5) is rapidly transformed into a linear relation \cite{4}:
$\Delta_0^\ast \approx \Delta_0(\Delta_s/ \omega_0)^{\eta^2/4}$,
renormalization of $C$ vanishes, and $\Gamma_s(T)$ below $T_c$ may increase
before ``freezing out''.

\begin{figure}[tb]
\epsfxsize=7.5cm\centerline{\epsffile{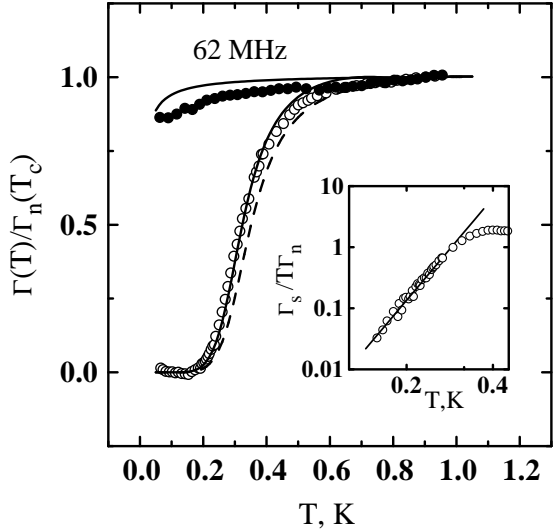}} \vspace{0.1in}
\caption{Temperature dependence of attenuation in $n$ ($\bullet$)
and $s$ ($\circ$) states. The solid curves correspond to
calculations for $\eta = 0.65$ and $T_c = 0.83$~K and the dotted
one to $T_c = 0.9$~K. The inset illustrates the evaluation of
$\Delta_s(0)$ and $\eta$. The solid line corresponds to the linear
approximation.} \label{fig1}
\end{figure}

Let us now discuss the experimental results. Figure~\ref{fig1} shows the
dependence $\Gamma_s(T)/ \Gamma_n(T_c)$ for transverse sound. The
normalization factor $\Gamma_n(T_c)^{-1}$ used for presenting the results can
be easily determined from the variation of the amplitude of the acoustic
signal between $T_c$ and the deep superconducting state.

In the region of the low-temperature ``tail'', the renormalization $g(u)$ can
be disregarded. The following estimate is obtained from (1), (2) and (4):
\begin{equation}
{\Gamma_s(T)\over \Gamma_n(T_c)}={2\pi \eta^2 \over 3\omega}
Te^{-\Delta_s/T}, \;\;(T/T_c < 0,3).
\end{equation}

According to this equation, the low-temperature region $\Gamma_s(T)$ must
become linear in coordinates $\ln(\Gamma_s(T)T^{-1})$, $T^{-1}$, as it can be
seen from the inset to Fig.~1. The slope of the approximating straight line
is determined by the superconducting energy gap which is in good accord with
the BCS value: $\Delta_s(0)/T_c = 1.7\pm 0.1$ ($T_{c}=0,9$ K). Intersection
of the approximating straight line with the ordinate axis leads to the
estimate $\eta = 0.65 \pm 0.05$. These data can also be used to refine the
value of $T_c$ for the bulk of the sample. The slope of the approximating
straight line and its interaction with the ordinate axis (see inset to
Fig.~1) are by no means connected with the choice of $T_c$. For a complete
evaluation of $\Gamma_s(T)$, we must use the specific value of $T_c$. It can
be seen from Fig.~1 that the theoretical curve for $T_c = 0.83$~K is in much
better accord with the experimental data.

The closeness of $\Delta_s(0)/T_c$ to the BCS value indicates that
peculiarities in the behavior of $\Gamma(T)$ in the vicinity of $T_c$ are not
connected with the magnetic depairing effects. Let us now discuss the
applicability of the hypothesis of the electron renormalization of $C$ to the
description of the behavior of $\Gamma_s(T)$.

\begin{figure}[tb]
\epsfxsize=7.5cm\centerline{\epsffile{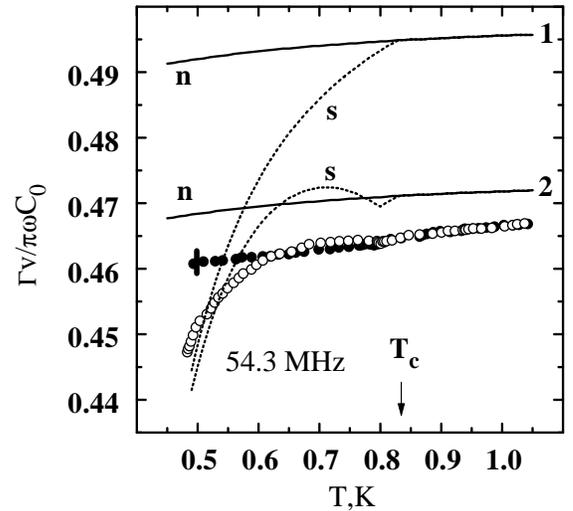}} \vspace{0.1in}
\caption{Behavior of attenuation in the vicinity of $T_c$ in $n$
($\bullet$) and $s$ ($\circ$) states. The experimental recordings
are averaged, and the noise level is indicated by the bold
vertical line. Curves~{\em 1} and {\em 2} are calculated without
and with the account for renormalization respectively; $\eta =
0.65, d = 1.2$.} \label{fig2}
\end{figure}

Figure~\ref{fig2} shows the data on variation of $\Gamma_s(T)$ and
$\Gamma_n(T)$ in the vicinity of $T_c$, measured with a higher resolution
than in Fig.~1. The results are normalized to the value $C_0 = 2.85 \cdot
10^{-5}$ obtained from the slope of $v_s(\ln T)$ in the deep superconducting
state ($T \lesssim 0.3$ K).$^{1)}$ The normalized value of $\Gamma(T_c)$ in
the standard TM must be close to 0.5 for $\omega \ll T$. Renormalization
(decrease) of $C$ naturally shifts $\Gamma$ towards lower values. However, it
has not been possible to measure the value of attenuation with an accuracy
better than $1\%$, which would allow an analysis of the shift of the
experimental dependence relative to the theoretical one. Hence we consider
only relative position of the lines $\Gamma_s(T)$ and $\Gamma_n(T)$ (the
latter value is obtained in a magnetic field $H \simeq 2$~T) which could be
measured with a much higher accuracy.

The meaning of the ``anomaly'' in $\Gamma$ discussed above can be seen
clearly from a comparison of the experimental dependences with the
theoretical ones obtained in the standard TM (curves~{\em 1} in Fig.~2).
According to calculations, a fall in absorption begins at $T_c$ and continues
with growing steepness upon a decrease in temperature. The experimental
dependence clearly displays a different behavior: $\Gamma_s(T)$ does not
display any variation at $T_c$ within the limits of resolution, and a
tendency towards an increase in $\Gamma_s(T)$ over $\Gamma_n(T)$ is observed
at lower temperatures. In any case, $\Gamma_s(T)$ remains practically
unchanged over a wide temperature range $T < T_c$. More clearly manifested
effects of this kind were detected earlier in the alloy Pd$_{30}$Zr$_{70}$
\cite{3}. It was also suggested in Ref.~3 that these anomalies can
be associated with the electron renormalization of the parameter $C$.

The renormalization of $C$ indeed takes place in the alloy investigated by
us. An irrefutable proof of this is the intersection the dependences $v_s(T)$
and $v_n(T)$ at a quite low temperature $T_{\rm cr}$ (its value for 62~MHz is
0.055~K). The scale of renormalization is quite significant ($\delta C/C_0
\simeq 0.25$) and is about double the quantity $\eta^2/4 \simeq 0.09$--0.12,
which allows us to assume the existence of several mechanisms of
renormalization.$^{2)}$ Moreover, incompatibility of the scale of $\delta
C/C_0$ with the anomalies in $\Gamma_s(T)$ indicates that these mechanisms
affect only insignificantly the TLS forming the relaxation attenuation for $T
\sim T_c$. It should be recalled that the main contribution to $\Gamma(T)$
comes from asymmetric TLS with $u_{\rm opt} \sim \sqrt{\omega/T} \ll 1$. One
possible mechanism of renormalization, which takes into account the
fluctuational rearrangement of the barrier in a double-well potential, is
associated only with the symmetric TLS \cite{8} and apparently makes no
contribution to $\Gamma(T)$.

The adiabatic renormalization does not impose any constraints on the possible
values of $u$ \cite{7}. In spite of the fact that the condition $\varepsilon
> 1$ moves the coherently tunneling TLS to the region of action of the
truncating factor in~(2), their partial contribution to $\Gamma(T)$ may be
quite significant on the scale of Fig.~2.

For the purpose of numerical computations, we used the model energy
dependence of the parameter $C$
\begin{equation}
C/C_0=1\!-\! \Theta(\varepsilon\!-\!d) \left[1\!-\!(2f(\Delta)\! -\!1)
\Theta(2\Delta\!-\! \varepsilon)\right] \eta^2/4,
\end{equation}
where $d$ is a fitting parameter. The first cofactor in the term describing
renormalization in Eq.~(8) defines the boundary between the regions~1 and 2.
Since the coherent amplitude $\Delta_0^*$ decreases exponentially for
$\varepsilon < 1$ \cite{4}, such an approximation seems to be reasonable. The
second cofactor in (8) takes into account that only normal excitations can
contribute to renormalization  for $\varepsilon < 2 \Delta$.

The results of calculations are also presented in Fig.~2 (curves~{\em 2}).
The interval of approximate ``independence'' of $\Gamma_s(T)$ can be matched
with that observed for a given value $\eta = 0.65$ for a quite reasonable
value of $d = 1.2 \pm 0.1$. It can be seen that the calculated dependence
$\Gamma_s(T)$ varies initially below $T_c$ in accordance with the standard
TM. Subsequently, $\Gamma_s(T)$ displays a kink with a reversal of the sign
of $d\Gamma/dT$ at $T = 2 \Delta_s/d$. The emergence of the kink is a
consequence of the use of a step approximation in~(8). As long as $2\Delta_s$
does not exceed the value of $E = Td$, superconductivity does not have any
effect on renormalization. Apparently, the restriction on the renormalization
of $C$ imposed from below by a smooth function of energy decreases the
variation of $\Gamma_s(T)$ in the vicinity of $T_c$ and eliminates the kink.
The same result is also arrived at by the broadening of the superconducting
transition which is quite natural for an amorphous sample. Hence it can be
really expected that $\Gamma_s(T)$ will not change at $T_c$, as is indeed
observed in the experiments.

Thus, the evolution of $\Gamma_s(T)$ near $T_c$ is determined by two factors,
viz., a drop in $\Gamma_s(T)$ due to a decrease in the relaxation rate $\nu$,
and an increase in $\Gamma_s(T)$ as a result of ``freezing out'' of the
renormalization of $C$. In contrast to the latter factor, the former is
frequency-dependent, and hence the resulting variation of $\Gamma$ will also
depend on frequency. Upon a decrease in $\omega$, the temperature interval in
which $\Gamma_s(T) > \Gamma_n(T)$ must expand, and vice versa. In particular,
calculations show that an increase in frequency by an order of magnitude
(measurements were made just at these frequencies by Esquinazi et al.
\cite{3}) completely masks the effect of the second factor for the same
values of $\eta, T_c$, and $d$. However, Esquinazi et al. \cite{3} carried
out measurements on glass with $T_c \simeq 2.5$ K. In this region,
$\nu$ is determined mainly by phonons and depends weakly on the state of the
electron subsystem. Under these conditions, the ``freezing out'' of
renormalization must give an even more pronounced effect than in our
experiments, as was apparently observed in \cite{3}.

In conclusion, let us formulate the main results. The experimental dependence
of the absorption of sound in the amorphous superconducting alloy
Zr$_{41.2}$Ti$_{13.8}$Cu$_{12.5}$Ni$_{10}$Be$_{22.5}$ was used to determine
the superconducting energy gap (which is found to be practically identical to
the gap obtained in the BCS theory) and the parameter $\eta$ characterizing
the intensity of interaction of TLS with electrons. The departures from the
predictions of the standard tunneling model observed in the vicinity of $T_c$
can be explained qualitatively and quantitatively by the adiabatic
renormalization of the coherent tunneling amplitude.

The authors are obliged to Prof. G. Weiss for drawing their attention to the
publications by Kagan and Prokof'ev \cite{4} and Stockburger et al. \cite{7}.

This research was partially supported by the State Foundation on Fundamental
Research in Ukraine (grant No. 2.4/153) and by Deutsche
Forschungsgemeinschaft via SFB 252. One of the authors (W. L. J.) wishes to
thank the US Energy Department  for grant (No. DE-FG03-86ER45242), while S.
V. Zh. is grateful to the von Humboldt Foundation for financial support.

$^1${\small This value of $C_0$ is double the analogous value presented in
Ref.~1. The departure is due to the fact that the value of $C_0$ was
estimated in Ref.~1 by using the linear dependence $v_n(\ln T)$  whose slope
depends significantly on the renormalization of $C$. The latter was not taken
into consideration in Ref.~1. The value of $\eta$ presented in Ref.~1 is also
found to be overstated for the same reason.}

$^2${\small The effect of renormalization of $C$ on the velocity of sound
will be considered in a separate publication.}


\begin{references}
\vspace{-1.7cm}
\bibitem{1}A. L. Gaiduk, E. V. Bezuglyi, V. D. Fil and W. L.
Johnson, Low Temp. Phys. {\bf 23}, 857 (1997).
\bibitem{2}S. Hunklinger and A. K. Raychaudhuri in: {\em Progress in
Low Temperature Physics} (Ed. by D. F. Brewer), Vol.~9, North-Holland,
Amsterdam (1986).
\bibitem{3}P. Esquinazi, H.-M. Ritter, H. Neckel, et al., Z. Phys.
B: Cond. Mat. {\bf 64}, 81 (1986).
\bibitem{4}Yu. Kagan and N. V. Prokof'ev, Zh. Eksp. Teor. Fiz. {\bf
97}, 1698 (1990) [Sov. Phys.: JETP {\bf 70}, 957 (1990)].
\bibitem{5}J. L. Black, B. L. Gyorffy, and J. J\"{a}ckle, Phil.
Mag. {\bf B40}, 331 (1979).
\bibitem{6}J. L. Black and P. Fulde, Phys. Rev. Lett. {\bf 43}, 453
(1979).
\bibitem{7}J. Stockburger, U. Weiss, and R. G\"{o}rlich, Z. Phys.
B: Cond. Mat. {\bf 84}, 457 (1991).
\bibitem{8}K. Vladar and A. Zawadowski, Phys. Rev. {\bf B28}, 1564,
1582, 1696 (1983).
\end{references}
\end{document}